\documentstyle[12pt]{JHEP3}
\newtheorem{theorem}{Theorem}
\newtheorem{proposition}{Proposition}

\title{Spectral Functions for Gauge Fields in Rindler-Like Spaces}

\author{Andrey A. Bytsenko \\
       Depto. de F\'{\i}sica, Univ. Estadual de Londrina, Paran\'a, Brazil\\
       E-mail: abyts@uel.br}

\author{Antonio Edson Gon\c calves\\
       Depto. de F\'{\i}sica, Univ. Estadual de Londrina, Paran\'a, Brazil\\
       E-mail: goncalve@uel.br}

\author{Vanderlei dos Santos Mendes\\
       Depto. de F\'{\i}sica, Univ. Estadual de Londrina, Paran\'a, Brazil\\
       E-mail: vsmendes@uel.br}

\abstract{ A class of conformal deformations of Rindler-like
spaces is analyzed. We study the spectral properties of the
Laplace operators associated with $p-$forms and acting in these
spaces and in their spatial sections. The spectral density of
continuum spectrum and the spectral zeta functions related to the
abelian $p-$forms in real compact hyperbolic manifolds are
obtained.}

\begin{document}

\section{Introduction}
In this paper we consider a class of conformal deformations of
Rindler-like spaces, whose spatial sections have metric
conformally related to the metric of hyperbolic spaces, and the
spectral properties of the Laplace operators acting on these
spaces. The real hyperbolic spaces play important role in
supergravity \cite{Duff}, superstring theory
\cite{Bytsenko2,Maldacena} and cosmology \cite{Aurich,Kaloper}.
The finite temperature effects for massive scalar fields in
Rindler spaces, in its conformal connection to hyperbolic spaces,
have been considered in \cite{Bytsenko0}. Here we analize gauge
theories based on abelian $p-$forms and calculate in the Rindler
case (non-compact manifold) the spectral datas of Laplacians and
the measure density useful for the trace of tensor kernels.
The spectral zeta functions of real compact hyperbolic spaces are
calculated explicitly using the Fried trace formula.

\section{Spectral functions of space forms
$M={\mathbb R}^+\times{X}^{N}$}

Let $M$ be a $D=(N+2)-$dimensional space and let
$
ds^2=g_{00}({\bf x})(dx^0)^2+g_{ij}({\bf x})dx^idx^j\:,
\, {\bf x}=\{x^j\}\,\,\,\, i,j=1,...,N+1 \:.
$
For the class of conformal deformations of the Riemannian metric
$g_{\mu \nu}$ the following relation holds:
\begin{equation}
\widetilde{g}_{\mu\nu}({\bf x})=e^{2\sigma({\bf x})}g_{\mu\nu}
({\bf x})\,, \,\,\,\,\, \sigma({\bf x})\in C^{\infty}(M)
\mbox{.}
\label{conformal}
\end{equation}
Let us consider static spaces admiting canonical horizons and
having the topology of the form
${\mathbb R}\times{\mathbb R}^{+}\times X^{N}$. The metric reads
\begin{equation}
ds^2=- b^2x^2dx_0^2+dx^2+d\Omega^2_{N}
\mbox{,}
\label{metric1}
\end{equation}
where $b$ is a constant factor, and
$d\Omega^2_{N}$ is the spatial metric related to the
$N-$dimensional manifold $X^{N}$.
If $X^{N}\equiv{\mathbb R}^{N}$, one has to deal with the Rindler
space.
If $X^{N}\equiv S^{N}$, $b=(D-3)/(2R_H)$ and $R_H$ is the
Schwarzschild radius of a black hole, then one is dealing
with a space which approximates, near the horizon and in the
large mass limit, a $D-$dimensional black hole \cite{Bytsenko0}.
In the Euclidean sector the metric (\ref{metric1})
may be written in the conformally related form:
\begin{equation}
d{\widetilde s}^2=d\tau^2+ x^{-2}
\left( dx^2+d\Omega^2_{N-1} \right)
\:,
\label{metric2}
\end{equation}
where $\sigma({\bf x})=-(1/2){\rm log} g_{00}$,
$d\tau= ib dx^0$, $x^{-2}$ is the conformal factor.
Associated to the conformally deformed metric (\ref{metric2})
we get the quantities $dV = x^{-N-1}\,dx\,dV_{N}$, and
\begin{eqnarray}
{{L}}^{(N+1, {\widetilde g})} & = & -{\mathfrak L}^{(N+1,
{\widetilde g})}
-\rho_0^2+Cx^2\,,
\nonumber \\
{\mathfrak L}^{(N+1, {\widetilde g})} & = & -x^2\partial_x^2
+ (N-2)x\,\partial_x
+x^2{\mathfrak L}^{(N, g)}
\mbox{,}
\label{conformal}
\end{eqnarray}
where ${\mathfrak L}^{(N,g)}$ is the Laplace operator on
the manifold $X^{N}$, $dV_{N}$ its invariant
measure, $\rho_0 = (N-1)/2$, and the real constant $C$ depends on
the scalar curvature.

Let us search for the spectral properties of the elliptic operator
${{L}}^{(N+1, {\widetilde g})}$.

\begin{theorem} (A. A. Bytsenko, G. Cognola and S. Zerbini
\cite{Bytsenko0})\,\,\,
For any suitable function $F(L^{(N+1, {\widetilde g})})$ the
following formulas hold:
\begin{eqnarray}
&& {\rm Tr} F\left(L^{(N+1, {\widetilde g})}\right) =
\int_0^\infty F(s^2)\,\mu_I(s)\,ds\,,
\\
&& \mu_I(s) =  \mu(s)\,\int_0^\infty
\sum_\alpha |K_{is}(x\lambda_\alpha)|^2 x^{-1}dx
\:,
\label{PMI}
\end{eqnarray}
where $K_{is}(y)$ are the Bessel functions of imaginary argument
and $\{\lambda_{\alpha}\}$ is the set of eigenvalues of the operator
$L^{(N+1, {\widetilde g})}$.
\end{theorem}
Here we derive a general expression for $\mu_I(r)$ by making use of
Eq.~(\ref{PMI}).

\begin{proposition}
Let ${\mathcal K}_t = \exp(-tL^{(N+1, {\widetilde g})})$ be the
heat kernel of the operator $L^{(N+1, {\widetilde g})}$,
${\mathcal K}_{\ell}(L^{(N+1, {\widetilde g})})$ are the Seeley-De Witt
coefficients in the kernel expansion, and
$\varepsilon$ is a horizon cutoff parameter in integrating
over the space coordinates in Eq. (\ref{PMI}). Then we have:
\begin{eqnarray}
&& \mu_I(r) = \sum_{\ell=0}^{\left[\frac{N-1}2\right]}
\frac{{\mathcal K}_{2\ell}({\mathfrak L}_p^{(N, {\widetilde g})})\,
\Phi(r, N+1-2\ell)}{N-2\ell}
(4\pi \varepsilon^{-2})^{(N-2\ell)/2}
+ \frac{1}{2\pi}\zeta(0|{\mathfrak L}^{(N, g)})
\nonumber \\
&& \times
\left[\frac{d}{ds}{\rm log}\zeta(s|{\mathfrak L}^{(N, g)})|_{s=0}
+ \psi(ir)+\psi(-ir)-
2{\rm log}\left(\frac{\varepsilon}{2}\right)-\pi\delta(r)\right],
\label{muIrFF}
\end{eqnarray}
where
\begin{equation}
\Phi(r, N+1) =
\mu(r){\rm Vol}(S^{N-1})\left(\frac{x}{2\pi}\right)^{N}
\int_0^{\infty}|K_{ir}(xs)|^2s^{N-1}ds\,,
\end{equation}
${\rm Vol}(S^{N-1})$ is the volume of the $(N-1)-$dimensional
sphere,
$\left[\frac{N-1}{2}\right]$ is the integer part of the number
$(N-1)/2$,
$\psi(z)$ is the logarithmic derivative of $\Gamma-$function
and $\delta(r)$ is the Dirac delta function.
For odd $N$, $\zeta(0|{\mathfrak L}^{(N, g)})$ is vanishing and so
the last term in Eq. (\ref{muIrFF}) disappears.
\end{proposition}
{\bf Proof}.
We can use the Mellin-Barnes representation for the Bessel function
of imaginary argument \cite{Gradshteyn}
$
K_{ir}^2(x\lambda_\alpha)= (-16\pi)^{-1/2}
\int_{\Re s>1}
\Gamma(s+ir)\Gamma(s-ir)\Gamma(s)[\Gamma(s+1/2)]^{-1}
(x\lambda_\alpha)^{-2s}\,ds,
$
and observe that, for $\Re s> N/2$, the sum over
$\alpha$ gives
$
\sum_\alpha\lambda_\alpha^{-2s}=\zeta(s|{\mathfrak L}^{(N,g)}).
$
Thus,
\begin{equation}
\mu_I(r)=\frac{\mu(r)}{8\sqrt{-\pi}}
\int_{\Re s=c> N/2}
\frac{\Gamma(s+ir)\Gamma(s-ir)\Gamma(s)
\zeta(s|{\mathfrak L}^{(N-1, g)})}
{s\Gamma(s+1/2)\varepsilon^{2s}}\,ds
\mbox{.}
\label{muIr}
\end{equation}
To make the integral we consider the rectangular contour
${\mathcal C}:=\{\Re s=c,\Im s=a,\Re s=-c,\Im s=-a\}$ and observe
that the two horizontal paths $\Im s=\pm a$ give a vanishing
contribution in the limit $a\to\infty$, as well as the path $\Re
s=-c$ in the limit $\varepsilon\to 0$. Also the poles in the strip
$-c<\Re s<0$ give a vanishing contribution as soon as
$\varepsilon\to 0$. Then we have to take into consideration only
the poles of the integrand in Eq.~(\ref{muIr}) in the half-plane
$\Re s\geq 0$. Such a function has simple poles at the points
$s=0$, $s=-n\pm ir$ and $s=(N-n)/2$ ($n\geq 0$). If $D$ is even,
that is $N$ is odd, $s=0$ is a double pole. It is clear that all
poles with $\Re s>0$ give rise to divergences, the number of them
depending on $N$, while the poles at $s=0$ and $s=\pm ir$ give
rise to finite contributions. Thus the result (\ref{muIrFF}) follows
from Eq. (\ref{muIr}). ${\Box}$

\section{The trace formula}

Let $\omega_p,\, \varphi_p$ be exterior differential $p-$forms; then,
the invariant inner product is defined by
$(\omega_p, \varphi_p)\stackrel{def}{=}\int_{X^N}
\omega_p\wedge*\varphi_p$. The following
properties for operators and forms hold: $dd=\delta\delta=0$,\, $\delta
= (-1)^{Np+N+1}*d*$,\, **$\omega_p = (-1)^{p(N-p)}\omega_p$.
The operators $d$ and $\delta$ are
adjoint to each other with respect to this inner product for
$p-$forms: $(\delta\omega_p, \varphi_p) = (\omega_p, d\varphi_p)$.
In quantum field theory the Lagrangian associated with $\omega_p$
takes the form: $d\omega_p\wedge *d\omega_p$ (gauge field)\,,\,
and $\delta\omega_p\wedge*\delta\omega_p$ (co-gauge field).
These Lagrangians provide a possible representation of tensor fields or
generalized abelian gauge fields. The two representations of
tensor fields are not completely independent, because of the
well-known duality property of exterior calculus which gives a
connection between star-conjugated gauge and co-gauge tensor
fields \cite{Obukhov}. The gauge $p-$forms are mapped into the co-gauge
$(N-p)-$forms under the action of the Hodge $*$ operator.

\begin{itemize}
\item{} The results (\ref{conformal}), (\ref{muIrFF}) can easy be
generalized for the case of the Laplacians ${\frak L}_p^{(N, g)}$
acting on $p-$forms.
\end{itemize}

Let $X_{\Gamma}=X^N$ be a $N-$dimensional real compact hyperbolic space
with universal covering $Y$ and fundamental group $\Gamma$.
Then we can represent $Y$ as the symmetric space $G/K$, where
$G=SO_1(N,1)$
and $K=SO(N)$ is a maximal compact subgroup of $G$. We regard
$\Gamma$ as
a discrete subgroup of $G$ acting isometrically on $Y$, and we take
$X_{\Gamma}$ to be the quotient space by that action:
$X_{\Gamma}=\Gamma\backslash Y= \Gamma\backslash G/K$.
Let $\tau$ be an irreducible representation of $K$ on a complex vector
space
$V_\tau$, and form the induced homogeneous vector bundle
$G\times_K V_\tau$
(the fiber product of $G$ with $V_\tau$ over $K$) over
$Y$. Restricting the $G$ action to $\Gamma$ we obtain the quotient
bundle
$E_\tau=\Gamma\backslash (G\times_KV_\tau)\rightarrow X_{\Gamma}$.
The natural Riemannian structure on $Y$ (therefore on $X_{\Gamma}$)
induced by the Killing form $(\;,\;)$ of $G$ gives rise to a connection
Laplacian ${\mathfrak L}_p^{(N, g)}$ on $E_\tau$.
Let $\sigma_p$ be the natural representation of $SO(2k-1)$
on $\Lambda^p {\mathbb C}^{2k-1}$,
and $\mu_{\sigma_p(r)}$ be the corresponding
Harish-Chandra-Plancherel density (given for a suitable
normalization of the Haar measure $dx$ on $G$).
Let ${\rm Vol}(\Gamma \backslash G)$ will denote the integral of the
constant function ${\bf 1}$ on $\Gamma \backslash G$ with respect
to the $G-$invariant measure on $\Gamma \backslash G$ induced by $dx$.
We can apply the version of the trace formula:

\begin{theorem} (D. Fried \cite{Fried})\,\,\,
For $0\leq p\leq N-1$ the trace formula applied to kernel
${\cal K}_{t} = e^{-t{\mathfrak L}_p^{(N,g)}}$ holds:
\begin{equation}
{\rm Tr}\left( e^{-t{\mathfrak L}_p^{(N,g)}}\right) =
I_{\Gamma }^{(p)}({\cal K}_{t})
+I_{\Gamma }^{(p-1)}({\cal K}_{t})
+ H_{\Gamma }^{(p)}({\cal K}_{t})
+H_{\Gamma }^{(p-1)}({\cal K}_{t})
\mbox{,}
\label{Fried}
\end{equation}
where $I_{\Gamma }^{(p)}({\cal K}_{t}),\, H_{\Gamma }^{(p)}({\cal
K}_{t})$ are the identity and hyperbolic orbital integrals
respectively. In the above formula
\begin{eqnarray}
I_{\Gamma }^{(p)}({\cal K}_{t}) & \stackrel{def}{=} &
\frac{\chi (1){\rm Vol}
(\Gamma \backslash G)}{4\pi }\int_{{\mathbb R}}\mu _{\sigma
_{p}}(r)
e^{-t(r^{2}+b^{\left( p\right) }+\left( \rho _{0}-p\right) ^{2})}dr,
\\
H_{\Gamma }^{(p)}({\cal K}_{t}) & \stackrel{def}{=} &
\frac{1}{\sqrt{4\pi t}}
\sum_{\gamma \in C_{\Gamma }-\{1\}}\frac{\chi (\gamma )}{j(\gamma )}
t_{\gamma }C(\gamma )\chi _{\sigma _{p}}(m_{\gamma })
e^{-t\left( b^{\left(
p\right) }+\left( \rho _{0}-p\right) ^{2}\right) - t_{\gamma }^{2}/4t}.
\label{H}
\end{eqnarray}
\end{theorem}
In Eq. (\ref{H}) $C_{\Gamma} \subset \Gamma$ is a complete set of
representations in $\Gamma$ of its conjugacy classes, $C(\gamma)$
is a well defined function on $\Gamma - \{1\}$ (for more details
see \cite{Bytsenko1,Bytsenko0,Williams,Bytsenko3}), $b^{(p)}$ are
real constants, and $\chi_\sigma(m)={\rm trace} (\sigma(m))$ is
the character $\sigma$ for $m\in SO(2k-1)$.

The spectral zeta function related to the Laplace operator
${\mathfrak L}_{p}^{(N, g)}$ can be represented by
the inverse Mellin transform of the heat kernel
${\mathcal K}_t$. Using the Fried formula, we can write the
zeta function as a sum of contributions:
\begin{eqnarray}
\zeta(s|{\mathfrak L}_p^{(N,g)}) & = &
\frac{1}{\Gamma (s)}
\int_{0}^{\infty}
\left(I_{\Gamma}^{(p)}({\mathcal K}_{t})+
I_{\Gamma }^{(p-1)}({\mathcal K}_{t})+H_{\Gamma }^{(p)}
({\mathcal K}_{t})+H_{\Gamma}^{(p-1)}({\mathcal K}_{t})
\right)t^{s-1}dt
\nonumber \\
& \equiv &
\zeta^{(N)}_{I}(s,p)+ \zeta^{(N)}_{I}(s,p-1)
+ \zeta^{(N)}_{H}(s,p) + \zeta^{(N)}_{H}(s,p-1)
\mbox{.}
\end{eqnarray}

\begin{proposition}\,\,\,
The identity and hyperbolic components of the spectral zeta function
can be presented in the form:
\begin{eqnarray}
\zeta_{I}^{(2k)}(s,p) & = & \frac{V_{\Gamma }C_{2k}^{(p)}}
{\Gamma(s)}\sum_{\ell =0}^{k-1}a_{2\ell ,2k}^{(p)}
\left[\frac{\Gamma (\ell +1)\Gamma (s-\ell -1)}
{\alpha_{p}^{2s-2\ell -2}}
+ \sum_{n=0}^{\infty }\frac{\xi_{n\ell }\Gamma (s+n)}
{\alpha_{p}^{2s+2n}} \right]\,,
\label{zeta2}
\\
\zeta _{I}^{(2k+1)}(s,p) & = & \frac{V_{\Gamma }C_{2k+1}^{(p)}}
{\Gamma (s)}\sum_{\ell =0}^{k}a_{2\ell ,2k+1}^{(p)}
\int_{0}^{\infty}t^{s-1}dt
\int_{\mathbb R}e^{-t(r^{2}+\alpha_{p}^{2})}r^{2\ell}dr
\nonumber \\
& = & \frac{V_{\Gamma }C_{2k+1}^{(p)}}
{\Gamma (s)}\sum_{\ell =0}^{k}
a_{2\ell, 2k+1}^{(p)}\Gamma \left( \ell+1/2\right)
\Gamma \left( s-\ell - 1/2\right)
\alpha_{p}^{-2s+2\ell +1}\,,
\label{zeta3}
\\
\zeta _{H}^{(N)}(s,p) & = & \!\! \sum_{\gamma \in C_{\Gamma}- \{ 1\}}
\frac{\chi (\gamma )t_{\gamma }^{2s}C(\gamma)
\chi _{\sigma_{p}}(m_{\gamma })}{\sqrt{\pi}\Gamma (s) j(\gamma )}
\frac{K_{-s+ \frac{1}{2}}(\alpha _{p}t_{\gamma })}
{(2\alpha t_{\gamma})^{s-1/2}}\,,
\label{zeta4}
\end{eqnarray}
where
$V_{\Gamma }=\chi (1){\rm Vol}\left(\Gamma \backslash G\right)/4\pi$,
and we have defined
$
\alpha_{p}^{2}:=b^{(p)}+\left( \rho_{0}-p\right)^{2}\,,
$
\begin{equation}
\xi_{n\ell}\stackrel{def}{=}  \frac{(-1)^{\ell +1}
\left(1-2^{-2\ell -2n-1}\right)}{n!(2\ell+2n+2)}B_{2\ell +2n+2}
\mbox{.}
\label{csi}
\end{equation}
\end{proposition}

{\bf Proof}. For the identity component we get
\begin{equation}
\zeta_{I}^{(N)}(s,p)=
\frac{V_{\Gamma}}{\Gamma (s)}
\int_{0}^{\infty }t^{s-1}dt\int_{{\mathbb R}}\mu_{\sigma _{p}}
e^{-t(r^{2}+\alpha_{p}^{2})}dr.
\end{equation}
For $\sigma_p$ the natural representation of $SO(2k-1)$
on $\Lambda^p {\mathbb C}^{2k-1}$, we have the corresponding
Harish-Chandra-Plancherel density given -- for a suitable
normalization of the Haar measure $dx$ on $G$ -- by
\begin{equation}
\label{07}
\mu_{\sigma_p(r)}= \frac{\pi}{2^{4k-4}[\Gamma(k)]^2} \left(
\begin{array}{c}
2k-1 \\
p
\end{array}
\right) P_{\sigma_p}(r)r \tanh(\pi r)\;,
\end{equation}
for $0\le p \le k-1$, where
\begin{equation}
\label{08}
P_{\sigma_p}(r) = \prod_{\ell=2}^{p+1} \left[ r^2+\left(k-\ell+\frac{3}{2}
\right)^2 \right]
\prod_{\ell=p+2}^{k} \left[ r^2+\left(k-\ell+\frac{1}{2}
\right)^2 \right]\;
\end{equation}
is an even polynomial of degree $2k-2$. We have that $P_{\sigma_p}(r)=
P_{\sigma_{2k-1-p}}(r)$ and
$\mu_{\sigma_p}(r)=\mu_{\sigma_{2k-1-p}}(r)$ for
$k\le p\le 2k-1$. Define the Miatello coefficients
\cite{Miatello,Bytsenko4}
$a_{2\ell}^{(p)}$ for $G=SO_1(2k+1, 1)$ by
$
P_{\sigma _{p}}(r)=\sum_{\ell =0}^{k-1}a_{2\ell }^{(p)}r^{2\ell },
$
$
0\leq p\leq 2k-1.
$
Replacing the Harish-Chandra-Plancherel measure, we obtain two
representations for $\zeta_{I}^{(N)}(s,p)$, which
holds for the cases of odd and even dimension.
Using the identities
$
{\rm tanh}(\pi r) = 1- 2(1+e^{2\pi r})^{-1},
$
and
$
\int_{0}^{\infty }(1+e^{2\pi r})^{-1}r^{2\ell -1}dr
=(-1)^{\ell -1}(1-2^{1-2\ell})(4\ell)^{-1}B_{2\ell},
$
where $B_{\ell}$ is the $\ell-$th Bernoulli number, we get
Eqs. (\ref{zeta2}) and (\ref{zeta3}).
Finally using the following representation for the Bessel
function \cite{Gradshteyn}
$
K_{\nu}(z) = (1/2)(z/2)^{\nu}
\int_{0}^{\infty}e^{-t-\frac{z^{2}}{4t}}t^{-\nu-1}dt
$
($|{\rm arg}\, z|<\pi/2$ and $\Re\, z^2>0$) we get formula (\ref{zeta4}).
${\Box}$



\begin{thebibliography}{99}

\bibitem{Duff}
M. J. Duff, B. E. Nilsson and C. N. Pope, {\it Kaluza-Klein
Supergravity, Phys. Rep.} {\bf 130} (1986) 1. \footnote{see also
the paper of A. A. Bytsenko, M. E. X. Guimar\~{a}es and J. A.
Helay\"{e}l-Neto: {\it Hyperbolic Space Forms and Orbifold
Compactification in M-theory}, which is published in this volume.}

\bibitem{Bytsenko2}
A. A. Bytsenko, G. Cognola, L. Vanzo and S. Zerbini,
{\it Quantum Fields and Extended Objects in Space-Times with Constant
Curvature Spatial Section, Phys. Rep.} {\bf 266} (1996) 1
[hep-th/9505061].

\bibitem{Maldacena}
J. Maldacena, {\it The Large N Limit of
Superconformal Field Theories and Supergravity, Adv. Theor. Math.
Phys.} {\bf 2} (1998) 231 [hep-th/9711200].

\bibitem{Aurich}
R. Aurich, S. Lustig, F. Steiner and H. Then,
{\it Hyperbolic Universes with a Horned Topology and the CMB
Anisotropy, Class. Quant. Grav.} {\bf 21} (2004) 4901 [astro-ph/0403597].

\bibitem{Kaloper}
N. Kaloper, J. March-Russell, G. D. Starkman and M.
Trodden, {\it Compact Hyperbolic Extra Dimensions: Branes, Kaluza-Klein
Modes and Cosmology, Phys. Rev. Lett.} {\bf 85} (2000) 928
[hep-ph/0002001].

\bibitem{Bytsenko0}
A. A. Bytsenko, G. Cognola and S. Zerbini,
{\it Finite Temperature Effects for Massive Fields in D-Dimensional
Rindler-Like Spaces, Nucl. Phys.} {\bf B458} (1996) 267 [hep-th/9508104].

\bibitem{Gradshteyn}
I. S. Gradshteyn and I. M. Ryzik, {\it Table of
Integrals, Series, and Products, Collected and Enlarged Edition
Prepared by A. Jeffrey}, Academic Press, New York, 1980.

\bibitem{Obukhov}
Yu. N. Obukhov, {\it The Geometrical Approach to
Antisymmetric Tensor Field Theory, Phys. Lett.} {\bf B109}
(1982) 195.

\bibitem{Fried}
D. Fried, {\it Analytic Torsion and Closed Geodesics on
Hyperbolic Manifolds, Invent. Math.} {\bf 84} (1986) 523.

\bibitem{Bytsenko1}
E. Elizalde, S. D. Odintsov, A. Romeo, A. A. Bytsenko
and S. Zerbini, {\it Zeta Regularization Techniques with Applications},
World Scientific, Singapore, 1994.

\bibitem{Williams}
F. L. Williams, {\it Topological Casimir Energy for a
General Class of Clifford-Klein Space-Times, J. Math. Phys.}
{\bf 38} (1997) 796.

\bibitem{Bytsenko3}
A. A. Bytsenko, G. Cognola, E. Elizalde, V. Moretti and
S. Zerbini, {\it Analytic Aspects of Quantum Fields}, World Scientific,
Singapore, 2003.

\bibitem{Miatello}
R. Miatello, {\it The Minakshisundaram-Pleijel Coefficients for
the Vector-Valued Heat Kernel on Compact Locally Symmetric
Spaces of Negative Curvature, Trans. Am. Math. Soc.} {\bf 260}
(1980) 1.

\bibitem{Bytsenko4}
A. A. Bytsenko, E. Elizalde and M. E. X. Guimar\~{a}es,
{\it Operator Product on Locally Symmetric Spaces of Rank One and the
Multiplicative Anomaly, Int. J. Mod. Phys.} {\bf A18 }(2003)
2179 [hep-th/0305031].



\end{thebibliography}
\end{document}